\documentstyle[12pt]{article}
\title{\bf
TRAJECTORY-COHERENT STATES AND THE HEISENBERG'S UNCERTAINTY RELATION}
\author{{\bf A.G. Karavayev \ }\\
{\it Tomsk Polytechnic University , 634004 Tomsk, Russia }}
\date{}
\begin{document}
\maketitle
\begin{abstract}
\noindent
In this paper we investigate the problem of minimization the
Heisenberg's uncertainty relation by the trajectory-coherent states.
The conditions of minimization for Hamiltonian and trajectory
are obtained. We show that the trajectory-coherent states minimize
the Heisenberg's uncertainty relation for special Cauchy problem
for the Schr\"{o}dinger equation only.
\end{abstract}

     Since  the   coherent  states   for the one-dimensional  harmonic
oscillator  were   first  constructed   by  Schr\"{o}dinger [1],they  have
been  widely  used   to  describe  many   fields  of  physics  [2].The
representation  of  coherent   states  has  been   used as a basis  in
almost  all   branches  of   physics  e.g. quantum  mechanics, quantum
optics, condensed matter   physics, atomic physics,   nuclear  physics
and mathematical physics.   At present the  coherent state   sequence
is  assumed  to   be  determined  as   a complete system of quantum  -
mechanical  states  which   are  eigenstates  of   the    annihilation
operators -  integrals  of  motion.    A quite  reasonable  conviction
has developed that coherent states are the closest to classical  in  a
certain   sense.   In   particular,   for   the quadratic  systems the
coordinate  and    momentum  quantum    -  mechanical    averages  are
solutions  of  the   classical   Hamiltonian   equations,  and    they
represent the minimum  uncertainty states in   the case of   quadratic
systems with constant coefficients.

     Recently  Bagrov,Belov   and  Ternov    [3]   have    constructed
approximate  (for  $  \hbar  \rightarrow  {  0}$  )  solutions  of the
Schr\"{o}dinger equation  for particles   in   general   potentials,  such
that  the coordinate and  momentum quantum - mechanical averages  were
exact  solutions    of   the  corresponding   classical    Hamiltonian
equations;   these  states   were  called  trajectory- coherent (TCS).
The  basis  of  this   construction  is  the  complex  WKB  method  by
V.P.Maslov [4-6].

    The  aim  of   this  work  is   to  investigate  the   problem  of
minimization  of   the  Heisenberg's   uncertainty relation by   TCS.
The conditions for  Hamiltonian and trajectory,  which guarantee    the
minimization  are   obtained.In  particular,we   show  that even   for
quadratic   system    (harmonic   oscillator)    TCS   minimize    the
Heisenberg's  uncertainty  relation    only  for   special     Cauchy
problem for the Schr\"{o}dinger equation.

     The Schr\"{o}dinger equation for the particles in general potentials
has the form
\begin{equation}
    i\hbar\partial_t\Psi=\hat{H}\Psi,                          
\end{equation}
where $ \hat{H}=(2m)^{-1}\hat{p}^2+V(x,t),~ \hat{p}=-i\hbar\partial_x$,~
and $m$   is the mass of the particle.

     The  method  of  construction  TCS  in  the  case  when symbol of
operator  $\hat{H}$  -  the  function  $H(x,p,t)$  was  arbitrary real
and  analytical  function  of   coordinate  and  momentum  has   been
described  in  detail  in  [3],hence  we  shall  only  illustrate some
moments.  For constructing the  TCS of the Schr\"{o}dinger equation  it is
necessary to solve the classical Hamiltonian system
\begin{equation}
   \dot{x}(t)=\partial_p{H(x,p,t)},~ \dot{p}(t)=-\partial_x{H(x,p,t)},
\end{equation}
and the system in variations (this is the linearization of the
Hamiltonian system in the neighbourhood of the trajectory $x(t),p(t)$)
\begin{eqnarray}
\dot{w}(t)=-H_{xp}(t)w(t)-H_{xx}(t)z(t),&   w(0)=b,   \\             
\dot{z}(t)=H_{pp}(t)w(t)+H_{px}(t)z(t), &   z(0)=1, \nonumber
\end{eqnarray}
where $H(x,p,t)$ is the classical Hamiltonian
\begin{eqnarray}
H_{xp}(t)= \partial_x\partial_p{H(x,p,t)} \mid_{x=x(t),p=p(t)},~
H_{px}(t)= \partial_p\partial_x{H(x,p,t)} \mid_{x=x(t),p=p(t)},\nonumber\\
H_{xx}(t)= \partial_{xx}^2 {H(x,p,t)} \mid_{x=x(t),p=p(t)},~
H_{pp}(t)= \partial_{pp}^2 {H(x,p,t)} \mid_{x=x(t),p=p(t)}, \nonumber
\end{eqnarray}
$b$ is complex number obeying the condition $ { \bf Im}b> 0$,~and
~$x(t),~p(t)$~are the solutions of system (2).

Obviously,the system (3) for the equation (1) is simpler,then
in general case:
\begin{eqnarray*}
\dot{w}(t)=-\partial_{xx}^2{V(x,t)} \mid_{x=x(t)}z(t),&w(0)=b,&{\bf Im}b> 0,\\
\dot{z}(t)=m^{-1}w(t),&z(0)=1.
\end{eqnarray*}

    The function of WKB - solution type [3]
\begin{equation}
\Psi_0(x,t,\hbar)=N\Phi(t)\exp\{i\hbar^{-1}S(x,t)\},                
\end{equation}
where $N=({\bf Im}b(\pi\hbar)^{-1})^{-1/4},~ \Phi(t)=(z(t))^{-1/2},$
\begin{eqnarray}
S(x,t)=\int\limits_{0}^{t}{\{\dot{x}(t)p(t)-H(x(t),p(t),t)\}dt}+ \nonumber
p(t)(x-x(t))+ \\+\frac{1}{2}w(t)z^{-1}(t)(x-x(t))^{2},\nonumber
\end{eqnarray}
and the phase $S(x,t)$ is the complex - valued function $({\bf Im}S> 0)$
is the approximate solution of the Cauchy problem for the Schr\"{o}dinger
equation with the initial value
\begin{eqnarray*}
\Psi_0(x,t,\hbar)\mid_{t=0}=N\exp\{i\hbar^{-1}(p_0(x-x_0)+\frac{b}{2}
(x-x_0)^2)\},
\end{eqnarray*}
where $x_0=x(t)\mid_{t=0},~p_0=p(t)\mid_{t=0}.$

We should note that if  $$\frac{\partial^k{V(x,t)}}{\partial{x^k}}
\mid_{x=x(t)}=0,\  k> 2, $$
then the function(4) is the exact solution of the Schr\"{o}dinger
equation (1) [7], for example, in the case of the quadratic systems
with constant coefficients.  We shall find the
$$(\Delta\hat{x})^2=<(\hat{x}-<\hat{x}>)^2>,\  (\Delta\hat{p})^2=
<(\hat{p}-<\hat{p}>)^2>,$$
where,for example,
$$<\hat{x}>=\int\Psi_0^*\hat{x}\Psi_odx.$$
Note, that
$$w(t)z^{-1}(t)-w^*(t)(z^{-1}(t))^*=2i{\bf Im}b\mid{z(t)}\mid^{-2},$$
because $w(t)z^*(t)-w^*(t)z(t)$ not depends on the time since  the
symplectic production of the solutions of the system in variations
(3) is equal to constant value.

By using this fact, we can obtain the following expressions for
$<\hat{x}>,\\ <\hat{p}>,\  (\Delta\hat{x})^2,\  (\Delta\hat{p})^2 $:
\begin{eqnarray}
<\hat{x}>=x(t),\ <\hat{p}>=p(t),\\               
(\Delta\hat{x})^2=\frac{\hbar}{2}\frac{\mid{z(t)}\mid^2}{{\bf Im}b},\
~(\Delta\hat{p})^2=\frac{\hbar}{2}\frac{\mid{w(t)}\mid^2}{{\bf Im}b}.
\nonumber
\end{eqnarray}

It follows from (5) that the Heisenberg's
uncertainty relation is minimized if: firstly, it must be\  ${\bf Re}b=0$,\
otherwise
$$
(\Delta\hat{x})^2(\Delta\hat{p})^2=\frac{\hbar^2}{4}\frac{({\bf Re}b)^2+
({\bf Im}b)^2}{({\bf Im}b)^2}>\frac{\hbar^2}{4}
$$
even at initial moment~($t_0=0$)~; secondly, the relationship\
$\mid{w(t)z(t)}\mid^2 =const$\  must satisfied for any\ $t$.  Obviously, the
    relationship\ $\mid{w(t)z(t)}\mid^2=const$\ is equivalent to equation
\begin{equation}
 \partial_t(w(t)z^*(t)w^*(t)z(t))=0,               
\end{equation}
and, by differentiating (6), taking into account (3) we obtain
\begin{equation}
H_{pp}(t)\mid{w(t)}\mid^2(w(t)z^*(t)+w^*(t)z(t))=    
H_{xx}(t)\mid{z(t)}\mid^2(w(t)z^*(t)+w^*(t)z(t)).
\end{equation}

The relationship (7) is satisfied in the following cases only:
firstly,when
$$w(t)z^*(t)+w^*(t)z(t)=0$$
or, secondly,
$$
H_{pp}(t)\mid{w(t)}\mid^2=H_{xx}(t)\mid{z(t)}\mid^2.
$$
Let us study the first case.

It is not difficult to show that the function\ $Q(t)=w(t)z^{-1}(t)$\  is the
solution of the equation
\begin{equation}
\dot{Q}(t)+H_{pp}(t)Q^2(t)+2H_{px}(t)Q(t)+H_{xx}(t)=0.    
\end{equation}

We denote\  $Q_1(t)={ \bf Re }Q(t)$\  and\  $Q_2(t)={ \bf Im }Q(t)$.\  It is
easy to show,that the equation\  $w(t)z^*(t)+w^*(t)z(t)=0$\ is equivalent
to\ $Q_1(t)=0.$\ Further,according to (8),\ $Q_1(t)$\ and\ $Q_2(t)$\
satisfy the systems of following equations:
\begin{equation}           
\dot{Q}_1(t)+H_{pp}(t)\{Q^2_1(t)-Q^2_2(t)\}+2H_{px}(t)Q_1(t)+H_{xx}(t)=0,
\end{equation}
\begin{equation}           
\dot{Q}_2(t)+2H_{pp}(t)Q_1(t)Q_2(t)+2H_{px}(t)Q_2(t)=0,
\end{equation}
$$
Q_1(0)=0,\  Q_2(0)={ \bf Im}b.
$$

Since\ $Q_1(t)=0$\  and,therefore,
$$
Q_2^2(t)=H_{xx}(t)H_{pp}^{-1}(t),
$$
we obtain from the equations (9),(10)
\begin{equation}
\partial_t\{H_{xx}(t)H_{pp}^{-1}(t)\}+4H_{px}(t)H_{xx}(t)H_{pp}^{-1}(t)=0.
\end{equation}                                                

For a particular case of\ $H_{px}(t)=0$\  (it satisfy for Eq.(1))
we have
\begin{equation}
H_{xx}(t)H_{pp}^{-1}(t)=const=({\bf Im}b)^2
\end{equation}                                                
and,therefore, the choice of the parameter\ $b$\  is not arbitrary.

Let us study the second case.
$$
H_{pp}(t)\mid{w(t)}\mid^2=H_{xx}(t)\mid{z(t)}\mid^2,
$$
and,taking into account the equalities
$$
w(t)z^{-1}(t)=Q(t),\   Q(t)=Q_1(t)+iQ_2(t),
$$
we obtain
$$
Q_1^2(t)+Q_2^2(t)=H_{xx}(t)H_{pp}^{-1}(t).
$$
Further, by using (9),(10) we have
\begin{equation}
\dot{Q}_1(t)+2H_{px}(t)Q_1(t)=-2H_{pp}(t)Q_1^2(t),\ Q_1(0)=0.  
\end{equation}

According to Cauchy theorem the Eq.(13) has the single solution
\ $Q_1(t)=0$\  and,therefore,\ ${ \bf Re }Q(t)=0$.\ As a result we have:
$$
w(t)z^*(t)+w^*(t)z(t)=0.
$$
Hence,in the second case we obtain as conditions for Hamiltonian
and trajectory as we have obtained in the first case (Eq.(11)-(12)).
For Hamiltonian\ $H(x,p,t)=(2m)^{-1}p^2+V(x,t)$\  the conditions
for minimization of the Heisenberg's uncertainty relation is defined by the
following statement:the minimization is possible provided the parameter\ $b$\
 and potential\ $V(x,t)$\  satisfy the following equations:
\begin{equation}
1)\ { \bf Re}b=0,~~~   \  2)~ \partial_{xx}^2V(x(t),t)=const=m^{-1}({ \bf
Im}b)^2 \end{equation}                                    
investigate now the Schr\"{o}dinger equation for the
potential~$V(x,t)=\frac{1}{2}m\omega ^2x^2$~(harmonic oscillator).
We obtain by similar calculations
\begin{eqnarray*}
x(t)=R\cos \omega t,~~~~~~p(t)=-m\omega R\sin \omega t,\\
w(t)=b\cos \omega t-m\omega \sin \omega t, ~~~~
z(t)=\frac {b}{m\omega }\sin \omega t+\cos \omega t.
\end{eqnarray*}
One may easily check by direct calculation that in this case wave
function (4) firstly,is the exact solution of Eq.(1) and,secondly
\ $<\hat{x}>=x(t),\ <\hat{p}>=p(t)$\  for any parameter\ $b$.\
But,according to (14) the equalities\ ${ \bf Re}b=0,\ { \bf Im}b=m\omega$\
must be satisfied.  Hence,we have
$$
w(t)=i\omega m\exp (i\omega t),\  z(t)=\exp (i\omega t),\ Q(t)=i\omega m.
$$
It is not difficult to show, that in this case TCS (4) minimize
the Heisenberg's uncertainty relation exactly.

\end{document}